\documentclass[12pt,preprint]{aastex}
\usepackage{amssymb}

\shorttitle{Line transfer in disks}
\shortauthors{Pavlyuchenkov et al.}

\begin{document}

\title{Molecular line radiative transfer in protoplanetary disks:\\
Monte Carlo simulations versus approximate methods}

\author{Ya. Pavlyuchenkov, D. Semenov, Th. Henning}
\affil{Max Planck Institute for Astronomy,
K\"onigstuhl 17, D-69117 Heidelberg, Germany}

\author{St. Guilloteau}
\affil{Laboratoire d'Astrophysique de Bordeaux, OASU, Universit\'e Bordeaux 1, CNRS, 2 rue de
l'Observatoire, BP 89, F-33270 Floirac, France}

\author{V. Pi{\'e}tu}
\affil{IRAM, 300 rue de la Piscine, F-38406
Saint Martin d'H\`eres, France}

\author{R. Launhardt}
\affil{Max Planck Institute for Astronomy,
K\"onigstuhl 17, D-69117 Heidelberg, Germany}

\and

\author{A. Dutrey}
\affil{Laboratoire d'Astrophysique de Bordeaux, OASU, Universit\'e Bordeaux 1, CNRS, 2 rue de
l'Observatoire, BP 89, F-33270 Floirac, France}

%
%

\begin{abstract}
We analyze the line radiative transfer in protoplanetary disks using several
approximate methods and a well-tested Accelerated Monte Carlo code. A low-mass
flaring disk model with uniform as well as stratified molecular abundances
is adopted. Radiative transfer in low and high rotational lines of CO,
C$^{18}$O, HCO$^+$, DCO$^+$, HCN, CS, and H$_2$CO is simulated. The
corresponding excitation temperatures, synthetic spectra, and channel maps
are derived and compared to the results of the Monte Carlo calculations.
A simple scheme that describes the conditions of the line excitation
for a chosen molecular transition is elaborated. We find that the simple
LTE approach can safely be applied for the low molecular transitions only,
while it significantly overestimates the intensities of the upper lines.
In contrast, the Full Escape Probability (FEP) approximation can safely be used for
the upper transitions ($J_{\rm up} \ga 3$) but it is not appropriate for the lowest transitions
because of the maser effect. In general, the molecular lines in protoplanetary
disks are partly subthermally excited and require more sophisticated
approximate line radiative transfer methods. We analyze a number of approximate
methods, namely, LVG, VEP (Vertical Escape Probability) and VOR (Vertical One Ray)
and discuss their algorithms in detail.
In addition, two modifications to the canonical Monte Carlo algorithm that
allow a significant speed up of the line radiative transfer modeling in rotating
configurations by a factor of 10--50 are described.
\end{abstract}


\section{Introduction}
The study of the formation and evolution of protoplanetary disks is an important
topic for our understanding of the formation of extrasolar planets and the
planets in our Solar system. The gas content of the disk that represents
most of its mass and has an imprint of earlier evolutionary history is best
studied in molecular lines. Since 1997, a number of rotational transitions
from simple molecules like CO, CS, CN, HCN, HCO$^+$, N$_2$H$^+$, H$_2$CO,
C$_2$H, and some of their isotopes have been firmly detected in several
nearby protoplanetary disks with single-dish telescopes and millimeter
interferometers \citep[e.g.,][]{Dutrey_etal1997,Kastner_etal1997,
Zadelhoff_etal2001,Aikawa_etal2003,Qi_etal2003,Qi_etal2004,Thi_etal2004,
Guilloteau_etal2006}. The wealth of information about disk physical structure
and chemical composition can only be acquired with multi-molecule,
multi-line interferometric observations at (sub-) millimeter wavelengths and at
sub-arcsecond resolution \citep[e.g.,][]{Qi_etal2006,Pietu_etal2007}.
In 2005, we started a comprehensive observational program at the IRAM 30-m
antenna and the Plateau de Bure interferometer, aimed at studying
protoplanetary disks with several key molecular tracers of disk temperature,
density, kinematics, and ionization degree
\citep[``Chemistry in Disks'' (CID) project; for the first results see][]{CID1}.

However, in order to extract all the relevant information from molecular lines
and thermal dust continuum one has to cope with the inverse (line) radiative
transfer (LRT) problem, which can hardly be solved for complex geometrical
configurations. Instead, one solves the direct LRT problem: models with varied
parameters are compared to the observational data until the best-fit is found
by using some iterative minimization technique \citep[e.g.][]{GD98,DDG03}.
Naturally, this task requires enormous amounts of computing power that severely
restricts the choice of applicable computational methods. This implies that
often fast but only approximate algorithms can be used. Thus it is crucial
to investigate the applicability of such approximate methods for the LRT modeling
of protoplanetary disks.

Various one- and multi-dimensional exact and approximate LRT methods have been
developed for modeling of stellar atmospheres, stellar winds, interstellar clouds,
and star-forming regions \citep[for a good overview see, e.g.,][]{Peraiah_2004}.
Among the simplest LRT concepts are the LTE (Local Thermodynamical Equilibrium)
and LVG (Large Velocity Gradients) approaches in which the LRT problem is treated
locally \citep[see][]{Sobolev1960,Mihalas1978}. The LVG method is just one among
the sub-class of ``Escape Probability'' approaches that allow the straightforward
calculation of the escape probability. These simplified LRT methods have been
applied to constrain the basic physical parameters of dense molecular cores using
single-dish molecular line data
\citep[see, e.g.,][]{deJong_etal1975,GLO81,Goldsmith_etal1983}.

Later more sophisticated, non-local LRT approaches have been developed, starting
from the 1D methods \citep[e.g.,][]{Lucas_1974,EAR06} to the multi-dimensional
(Accelerated) Monte Carlo, Local Linearization, and Multi-Zone Escape Probability methods, e.g.
\citet[][]{Bernes,Juvela_1997,Dullemond_Turolla2000,HvT00,Ossenkopf_etal2001,
Rawlings_Yates2001,Schoeier_Olofsson2001,Keto_etal2004,PS05}
\citep[for a recent review see][]{Zadelhoff_etal2002}. Being more accurate,
modern LRT methods are routinely
applied to study the kinematics and chemical structure of prestellar cores
in detail \citep[e.g.][]{Pavlyuchenkov_etal2006,Tafalla_etal2006}, but their use for protoplanetary
disks is still the exception \citep[e.g.][]{Semenov_etal2005,Bea07}.

The global appearance of the physical structure of protoplanetary disks
is commonly treated in the framework of steady-state passive or active
flaring disk models
with or without an inner rim, which seem to be successful in explaining
many observational facts
\citep[e.g.,][]{Dalessio_etal1998,Dalessio_etal2001,Dullemond_etal2001}.
The modeled structure can further be used to unravel the chemical evolution
of the disk with a sophisticated gas-grain chemical model including surface
processes, and even dynamical mixing and advective motions
\citep[e.g.,][]{Willacy_etal1998,Aikawa_etal2002,Gail2002, Ilgner_etal2004,
Semenov_etal2004,Semenov_etal2006,Willacy_etal2006,TG07}.
Finally, the disk physical structure and chemical composition together
with appropriate collision rates provide the necessary inputs for the
LRT simulations.

Since the iterative procedure to search for the best fit to the observed
spectra can easily become computationally prohibitive when the full
LRT has to be considered, it is of great importance to identify those fast
approximate LRT methods that give sufficiently accurate results (with respect to the
observational limitations) and to investigate the limits
of their applicability. The analysis is further complicated when a realistic
disk chemical structure is taken into account. To the best of our knowledge,
a detailed benchmarking of various approximate LRT approaches for protoplanetary disks
has not yet been attempted in the literature apart from the work by \citet{Zadelhoff_etal2001}
and therefore constitutes a major goal of the present paper.

In this paper, we compare a full 2D non-LTE Monte Carlo code with several
approximate line radiative transfer methods by considering a few representative
radiative transfer cases. These problems are thought to cover all possible
situations encountered in disks, namely, the excitation of optically thin/thick
emission lines in low/high density regions. Consequently, we consider low and
high transitions of key molecules in disks tracing temperature (CO and isotopologues),
column density ($^{13}$CO or C$^{18}$O), density (CS and H$_2$CO), deuterium fractionation
(DCO$^+$), ionization degree (HCO$^+$), and photochemistry (HCN). For this
exploratory study we adopt an irradiated flaring model of a young, low-mass
T~Tauri protoplanetary disk, and use both uniform molecular abundances
and those from an advanced chemical model (Section~\ref{model}). Our 2D Monte Carlo
algorithm is based on the 1D code ``RATRAN'' \citep{HvT00} with two acceleration
modifications for disks (see Section~\ref{method}). The adopted approximate
methods include the classical LTE approach, the Full Escape Probability (FEP)
and Vertical Escape Probability (VEP) methods, LVG, and a non-local 1D-method
VOR (Vertical One Ray method). The approximate methods are all described in
Section~\ref{approxi}. Using the input disk model and the above LRT methods,
in Section~\ref{results} we calculate the corresponding excitation temperatures,
synthetic spectra, and channel maps and compare these data to the results
obtained with the Monte Carlo code. The discussion and final conclusions follow.

\section{Disk physical and chemical structure}
\label{model}
\begin{figure}
\includegraphics[width=0.85\textwidth]{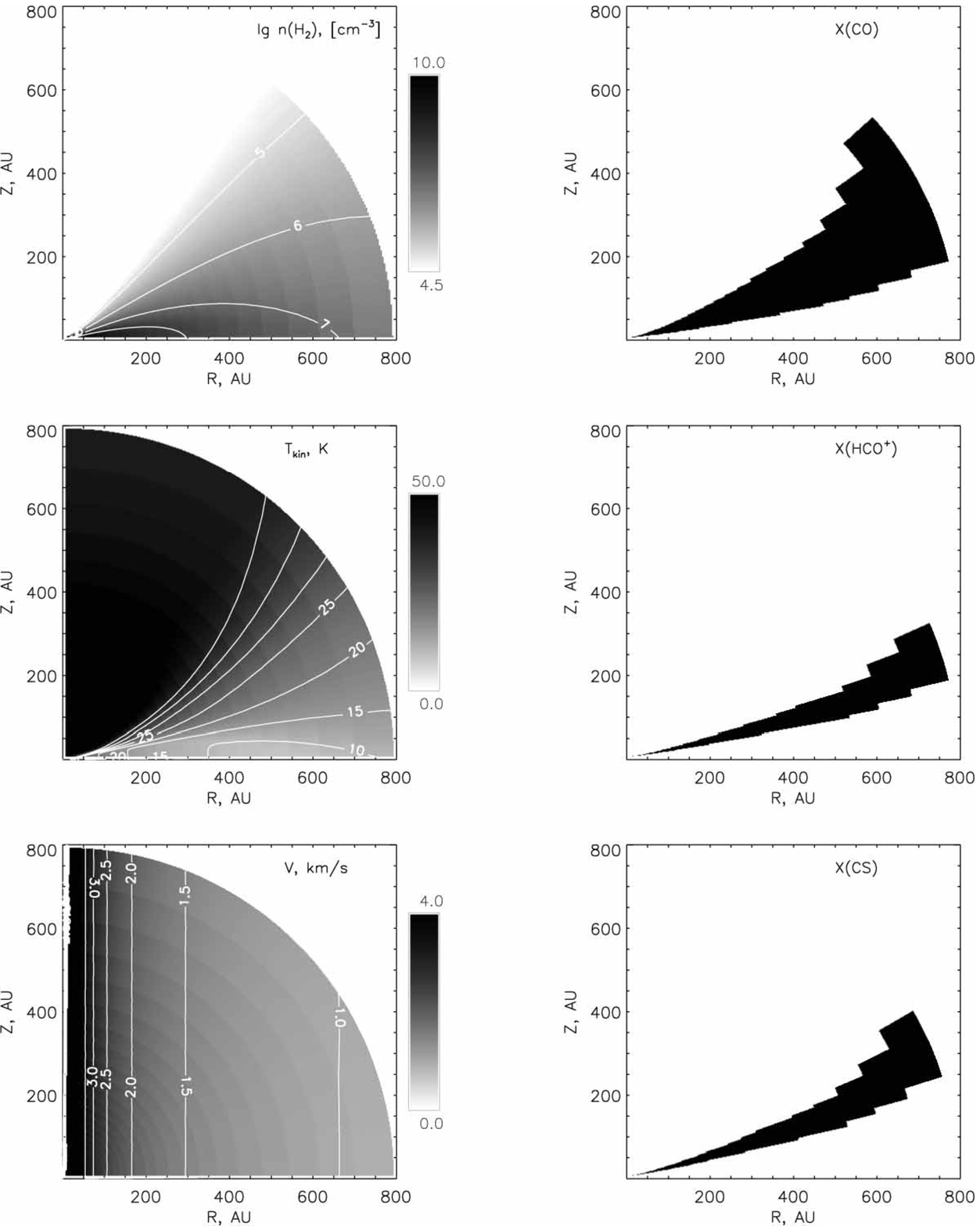}
\caption{Left: The distributions of density, temperature,
and radial velocity in the adopted disk model. Right: The
distributions of the molecular abundances relative to the
total amount of hydrogen nuclei for CO, HCO$^+$, and CS.}
\label{struct}
\end{figure}

We consider a typical protoplanetary disk surrounding a T Tauri-like star
in the nearby Taurus-Auriga molecular complex at 140~pc. A flaring steady-state
model representative of a Class II disk with vertical temperature gradient
is adopted \citep{Dalessio_etal1999}. The disk has inner and outer radii of 0.037
and 800~AU, respectively, a mass of $0.07\,M_{\sun}$, and a mass accretion
rate of $10^{-8}\,M_\sun$\,yr$^{-1}$. The central star has a radius of
$2.64\,R_{\sun}$, a mass of $0.7\,M_{\sun}$, an effective temperature
of 4\,000~K, and an age of about 5~Myr \citep[see Table~2 in][]{CID1}.
We assume that the disk has a constant microturbulent velocity
of 0.1~km\,s$^{-1}$ and a purely Keplerian rotation profile,
$V(r) \propto r^{-1/2}$ \citep{DDG03,Pietu_etal2007}. The disk structure
is shown in Fig.~\ref{struct} and the parameters are summarized
in Table~\ref{disk_param}.

In the present study we focus on the molecules that have been detected in
protoplanetary disks and are readily used as probes of the disk physical
and chemical structure, high-energy radiation, and deuterium fractionation:
CO, $^{13}$CO/C$^{18}$O, HCO$^+$, DCO$^+$, HCN, CS, and H$_2$CO.
Corresponding
to the different excitation conditions, the various rotational lines
of these molecules are thought to be generated in distinct disk regions ranging from
the cold midplane and warm molecular layer to the hot disk surface layer.

Normally one has to use a sophisticated gas-grain chemical network with
surface reactions in order to obtain realistic abundance distributions of
these observable species in the disk \citep[see][]{vanDishoeck_Blake1998,
Aikawa_Herbst1999,Markwick_ea02,Semenov_etal2005,Willacy_etal2006}.
For many molecules
their highest concentrations are reached in the disk intermediate layer
(at $\sim 0.5-1$ pressure scale heights) that is sufficiently opaque to
destructing high-energy radiation and yet too warm for effective freeze-out
of gas-phase molecules \citep[e.g.][]{Aikawa_etal2002,CID1}. As a result,
in these models, from a qualitative point of view the abundance distributions of many
molecules can be represented by a single layer with particular concentration,
thickness, and location in the disk \citep[see, however,][]{Semenov_etal2006,Aikawa_etal2007}.

However, for the first simple analysis of the observational data it is
often sufficient to assume that molecular abundances are constant
(with respect to hydrogen).
Moreover, the use of uniform molecular abundances can help to disentangle
most important line radiative transfer effects from strong chemical gradients
in the disks.

We adopt both approaches. The values of the uniform molecular concentrations
and parameters of the molecular layers are obtained by using the gas-grain
chemical model with surface reactions of \citet{Semenov_etal2005}. In essence,
this model is based on the UMIST\,95 database of gas-phase reactions
\citep{umist95}, a set of surface reactions from \citet{HHL92} and
\citet{HH93} on $0.1\mu$m silicate grains, and includes a small deuterium
network from \citet{BNM99}.

The location of each molecular layer in the disk is constrained by the upper
and lower boundaries that can be approximated by a power law:
$z(r)/H_p(r) \approx a \times r^b$ \citep[where $r$ is the radius in AU
and $H_p(r)$ is the pressure scale height as defined in][]{DDG03}.
We assume that the molecular abundances are constant across a molecular
layer and utilize the maximum values in the uniform disk model
(Fig.~\ref{struct}). Note that the CS abundances peak toward the inner disk
region and cannot be accurately represented by such a simple one-zone model.
The parameters of the uniform and layered chemical structures, i.e. parameters
$a_{min},b_{min}$ and $a_{max},b_{max}$ describing the upper and lower boundaries
as defined above, are presented in Table~\ref{chem_param}.

Finally, the collision rates for the line radiative transfer modeling
are taken from the publicly available ``Leiden Atomic and Molecular Database''
\footnote{\url{http://www.strw.leidenuniv.nl/$\sim$moldata}},
see \citet{Schoeier_etal2005}.

\section{Accelerated Monte Carlo method}
\label{method}
In this section we describe our implementation of the Accelerated Monte
Carlo technique in the 2D non-LTE line radiative transfer code ART,
which is part of the software package ``URAN(IA)'' aimed at modeling of
various objects with different exact/approximate LRT methods.

\subsection{Basic algorithm}
The general idea of all Monte Carlo radiative transfer methods is
to follow the propagation of randomly ejected photons or photon packages
through the medium. These codes are based on a straightforward formalism that
is easy to implement numerically and are capable of modeling radiative transfer
in various complex multidimensional objects with arbitrarily distributed sources
of radiation \citep[e.g.,][]{Juvela_1997,Wolf_etal1999,Pinte_etal2006,Nicollini2006}.
However, in order to lower random noise and thus retain reliable accuracy
of the results, Monte Carlo simulations typically require enormous amounts
of computing time, especially for highly optically thick objects. In certain
cases the numerically most demanding part of Monte Carlo codes (ray tracing)
can be significantly optimized by special iterative schemes for
the acceleration of convergence.

The algorithm of our 2D non-LTE line radiative transfer code ART is fully
described in \citet[][]{Urania} and only briefly summarized here. In essence,
this Monte Carlo code is based on the accelerated scheme originally proposed
and implemented in the LRT code ``RATRAN'' by \citet{HvT00} and in the modified
Monte Carlo method of \citet{Voronkov_1999}. The main equations of the line
radiative transfer can be found elsewhere
\citep[e.g.,][]{Zadelhoff_etal2002,Rohlfs_Wilson2004} and are not repeated here.

First, the computational domain is split into cells. All physical quantities are
assumed to be constant within each disk cell $i$, except for the velocity that can vary.
In the first iteration, initial level populations and a set of photon random directions
$\vec{n}(i)$ through the grid  are defined (this is the Monte Carlo part of the calculations).
The same set of photon random directions is used in all subsequent iterations for accurate
ray tracing, producing a solution free of random noise but with possible undersampling
of directions and velocities.

Using these values, the line intensities $I_{\nu}(i)$ are computed for each disk
cell $i$ by explicit direct integration of the radiative transfer equation with
the long characteristic method along the predefined set of photon rays $\vec{n}(i)$,
starting with the CMB field at the edge of the disk (or interstellar
radiation field, or anisotropic stellar radiation).
The integration path is split
into sub-intervals with a step that is smaller than the cell size. Ideally, one
would need to define this integration step based on local gradients of the physical
conditions in the disk and the optical depth of the line. However, this is not
easily achievable in practice, so an empirical value has to be used. We adopt
a step size of 1/10 of the cell size in our computations.

Next, the resulting mean line intensity $\tilde{J}$ is calculated in each disk
cell by spatial averaging of the derived intensities $I_{\nu}(i)$ over
directions $\vec{n}(i)$ and by averaging over the line profile. The computed
mean intensities are utilized in the next iteration step to refine the level
populations by solving the balance equations in all disk cells.

The convergence of the integration procedure for optically thick
lines can be substantially accelerated by additional internal sub-iterations
similar to the Accelerated Lambda Iterations (ALI) scheme \citep{Auer_Mihalas1969}.
The ALI method employs the fact that the calculated mean line intensity at each
particular grid cell can be separated into an internal component generated in the
cell itself and an external contribution coming from other grid cells. As a
result, additional sub-iterations are applied to bring into an agreement the
internal mean line intensity and the corresponding level populations in every
grid cell. When these local sub-iterations converge, the global iterations over
the entire grid are continued until the level populations in each cell do
not differ by more than $0.1$\%. Note that this scheme may not be fully appropriate
in cases of particularly thick lines with $\tau\ga 100$ when such a simple
convergence criterion fails, e.g. for water lines \citep{van_der_Tak_etal2004}.

After the final molecular level populations are obtained, the maximum random
error associated with the Monte Carlo computations is estimated. For that all
iterations are repeated again with another set of preselected random photon
rays. We choose the number of random photon paths such that the relative error
in the computed level populations is smaller than 1\%. Although the code is
currently restricted to axially symmetric (2D) radiative transfer problems
in polar coordinates, it can be extended to problems with any number of
dimensions and for any coordinate system.

Finally, these resulting level populations are used in the 3D ray tracing part
of the ``URAN(IA)'' package to produce synthetic molecular line spectra,
channel maps, integrated line intensity maps, velocity maps etc., with an
arbitrary number of velocity channels, and for any disk viewing geometry. Beam
convolution can be applied if needed.

The ``URAN(IA)'' code was carefully benchmarked against other LRT codes by
solving several one- and two-dimensional problems. In the one-dimensional regime,
the computed molecular level populations and excitation temperatures are fully
consistent with all tests proposed during the conference on molecular radiative
transfer\footnote{\url{http://www.strw.leidenuniv.nl/astrochem/radtrans/}}
\citep[Leiden University, 1999, see][]{Zadelhoff_etal2002}. Several additional
tests of our code were performed in 2D mode, using a few simple problems with
analytically predictable solutions. In addition, together with M. Hogerheijde
-- the developer of the two-dimensional LRT code ``RATRAN'' -- we successfully
compared the results for both codes for Keplerian disks.

\subsection{Further acceleration of the Monte Carlo algorithm}

\subsubsection{Concept of interaction areas}
The method of long characteristics is one of the most reliable algorithms to
calculate the mean intensity and thus to obtain an accurate solution to the
radiative transfer problem \citep[e.g.,][]{Dullemond_Turolla2000}.
However, a direct realization of this method is computationally expensive
for multi-dimensional radiative transfer simulations. For example, to compute
molecular level populations for a 2D disk model having 150 vertical and
50 radial grid points and 1000 photon packages per cell, one needs
$\ga5$~days of computational time on a modern PC (Pentium Xeon
2.4~GHz, 1Gb RAM).

However, for certain cases the computational time needed for the ray tracing
part can be efficiently reduced. To illustrate this for a rotating Keplerian disk
with radial velocity gradient, let us consider a cell at a given radius $r$.
The local line width that strongly depends on the temperature and microturbulent
velocity in the cell is typically $\sim 0.1-0.2$~km\,s$^{-1}$, which is
significantly smaller than the total velocity dispersion across the disk,
$V_\mathrm{disp} \ga 10$~km\,s$^{-1}$. Therefore, due to the Doppler shift the
photons emitted at a peak frequency in a cell rotating with a given velocity
can only be absorbed locally by some of the neighboring cells having similar
velocities (the so-called ``interaction area''). An example of such an interaction
area is shown in Fig.~\ref{interac}, where regions of the Keplerian disk that are
radiatively coupled to the cell located at $r \approx 130$~AU (white spot in the
figure) are shown in black.

We modified the ray-tracing algorithm used in the RT solver of ``URAN(IA)''
such that the iterative
integration of the RT equation is only performed over the radiatively coupled
regions of the disk. During the first iteration the radiatively coupled regions
for each cell are identified using the full ray tracing and stored in a file.
This information is utilized during the next iterations to integrate the RT
equation over the radiatively coupled regions only.
Note that the radiatively coupled region will be different for each line of each
of the considered molecules. However, for those molecules for which the local
lines widths are dominated by non-thermal motions (e.g., microturbulence) this
difference will be small.

This modification preserves the flexibility of the Accelerated Monte Carlo
approach and allows to decrease the computation time for protoplanetary disks
by about one order of magnitude and more, retaining the same high level of
accuracy. One may use an analytical prescription instead of the preliminary
ray tracing, but such a prescription will be model-dependent and thus cannot
be easily generalized. Though our concept of the interaction areas in rotating
configurations somewhat resembles the idea of the approximate LVG method
(see Sect.~\ref{LVG}), it leads to an \emph{exact} solution of the LRT problem.

\subsubsection{Concept of thermalized cells}
The densities in the disk interior are so high that the low level populations of
many molecules are thermalized. Naturally, the thermalized disk cells can be
effectively excluded from the global iteration calculations since the radiative
effects do not play a role in populating the levels. We use the following
approach to determine whether a disk cell is thermalized or not. First,
the level populations in the disk are calculated using the simple LTE and Full
Escape Probability (FEP) approaches (see next section). These two approximate
methods represent two extreme cases to the solution of the LRT problem.
The LTE approach tends to overestimate while FEP method tends to underestimate
the populations of high levels in the case of radiative coupling. The cell is
supposed to be thermalized if the relative difference between the corresponding
LTE and FEP level populations is smaller than an adopted threshold of
10$^{-3}$--10$^{-2}$. This modification leads to an additional computational
acceleration of about 10\% for HCO$^+$ and 30\% for CO and CO isotopes.

\section{Approximate LRT methods}
\label{approxi}
In this section, we summarize the approximate LRT methods that will be compared
with the results of the Monte Carlo simulations from the ART code. A schematic overview and
comparison of the utilized LRT approaches is given in Table~\ref{LRT_methods_table}.

\subsection{Local thermodynamical equilibrium}
\label{LTE}
In the case of local thermodynamical equilibrium (LTE), the molecular level
populations are thermalized. They follow a Boltzmann distribution with kinetic
gas temperature
$T_{kin}$:
\begin{equation}
\frac{n_i}{n_j} = \frac{g_i}{g_j}\exp\left\{-\frac{h\nu_{ij}}{kT_{kin}}\right\},
\label{eqlte}
\end{equation}
where $g_i$ and $g_j$ are the statistical weights and $\nu_{ij}$ is the frequency
of the $i-j$ transition. In terms of radiative transfer, thermalization means that
the excitation and kinetic temperatures are equal for a given transition:
\begin{equation}
T_\mathrm{kin}=T_\mathrm{exc}=\frac{h\nu_{ij}}{k}\left[ \ln \left( \frac{g_i}{g_j}\frac{n_j}{n_i} \right) \right]^{-1}.
\end{equation}
When the excitation temperature is higher or lower than the kinetic temperature,
the level populations are super- or subthermally excited.

There are two general situations when LTE holds: 1) The hydrogen density is
substantially higher than the critical density of a transition, or
2) the optical depth of a transition is so high that a perfect equilibrium
between the collisional and radiative (de-)excitations is reached. The LTE
approach is appropriate for the lower rotational transitions of CO and its isotopologues
(see \citet{DDG03,Pietu_etal2005}).
It has also been used in the analysis of molecular line data even if the
conditions for its applicability are not fully justified because it is
the easiest LRT method to implement \citep[e.g.,][]{Aikawa_etal2003,
Qi_etal2003,Narayanan_etal2006}.

\subsection{Full escape probability}
\label{FEP}
Another extreme is to assume that a transition has a very low optical depth,
so that the newly generated photons escape the disk freely, without any
further interaction with the medium.
Or, by other words, in the FEP approach the escape probability $\beta$ is
not calculated and always set to 1 \citep[see also][]{van_der_Tak_etal2007}.
Under such conditions the LRT problem
becomes local and can be easily solved. The Full Escape Probability (FEP)
level populations are obtained as a solution of the statistical equilibrium
equations with the mean intensity $J$ substituted by the mean external
radiation field, e.g. CMB: $J=J_\mathrm{CMB}$. As in the LTE case, the FEP
approach requires only very little computational time. FEP can be an
adequate method for molecules with very low abundances.

\subsection{Large velocity gradients}
\label{LVG}
In some objects such as expanding stellar envelopes or galaxies, radial
velocity gradients can be much larger than the local thermal and microturbulent
velocities \citep[e.g.,][]{Weiss_etal2005,Castro-Carrizo_etal2007}.
In this case the photons that are emitted by a certain region can only
be absorbed locally. The size of this region is determined by the local
velocity field and the line width. If one assumes that the physical conditions
in this local environment are homogeneous, the radiative transfer problem
can be substantially simplified \citep[e.g.][]{Sobolev1960,Castor1970,Goldreich_Kwan1974}.

\begin{figure}
\includegraphics[width=0.65\textwidth,clip=]{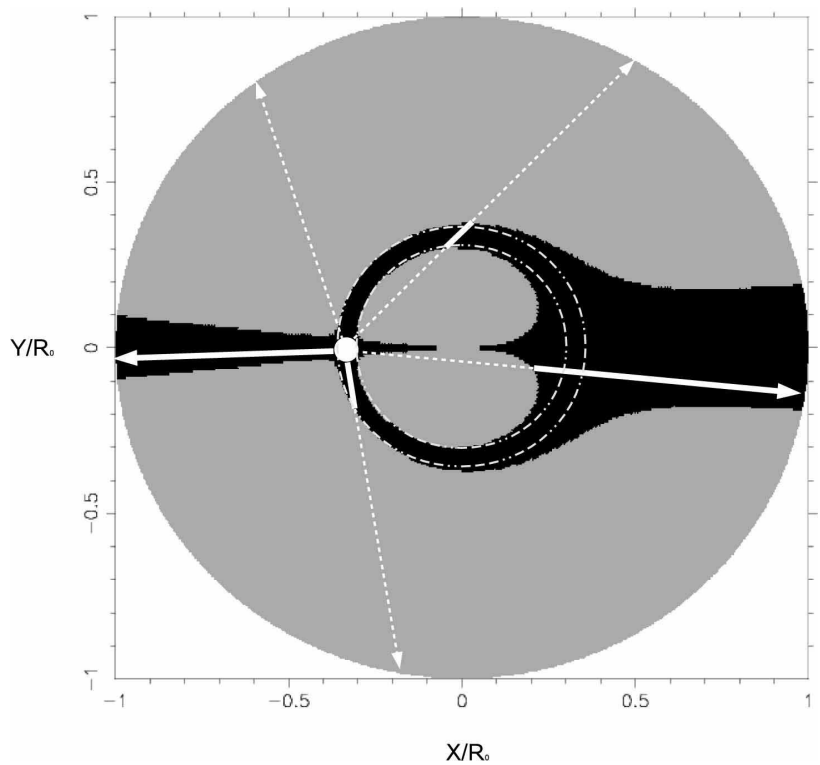}
\caption{The equatorial plane of a Keplerian disk with radius
$R_0$. The white spot represents a rotating cell that emits
radiation at a certain rest frequency. The regions where this
radiation can be absorbed by the same molecules are shown
in black (the so-called ``interaction area''). The interaction
area of the disk has velocities that are Doppler-shifted from
the velocity of the emitting cell by no more than the local
line width (expressed in km\,s$^{-1}$). The ring bounded by
white circles corresponds to the radiatively coupled region
in LVG method.}
\label{interac}
\end{figure}

The Keplerian rotation of a circumstellar disk causes strong velocity variations
of $\sim 0.5$--5~km\,s$^{-1}$, which are larger than the local thermal
($\sim 0.2$~km\,s$^{-1}$) and microturbulent ($\sim 0.2$~km\,s$^{-1}$)
velocities. As shown in Fig.~\ref{interac}, the size of the radiatively coupled
zone is significantly smaller than the entire disk plane. Therefore, one may
expect the LVG approach to be efficient for the line radiative transfer modeling.

In the LVG method, the mean intensity $J$ in a disk cell is approximated by a
combination of the source function $S$ for a given transition and the external
radiation field $J_\mathrm{ext}$:
\begin{equation}
J = (1-\beta)S+\beta J_\mathrm{ext},
\label{eqlvg1}
\end{equation}
where $\beta$ is the probability for the photons to freely escape the
medium, $0\le \beta \le 1$. The escape probability $\beta$ is usually
expressed in the form:
\begin{equation}
\beta=\frac{1-\exp(-\tau)}{\tau},
\label{eqlvg2}
\end{equation}
where $\tau$ is the effective line optical depth because of the local
velocity gradient. Note that for optically thin lines the escape probability
$\beta$ reaches unity and the LVG method turns into the FEP approach.

The effective optical depth $\tau$ is evaluated as:
\begin{equation}
\tau = \alpha\times\Delta L,
\end{equation}
where $\alpha$ is the local absorption coefficient and $\Delta L$ is the
coherence length i.e. the mean distance for a photon to escape the medium.

\begin{figure}
\includegraphics[width=0.45\textwidth]{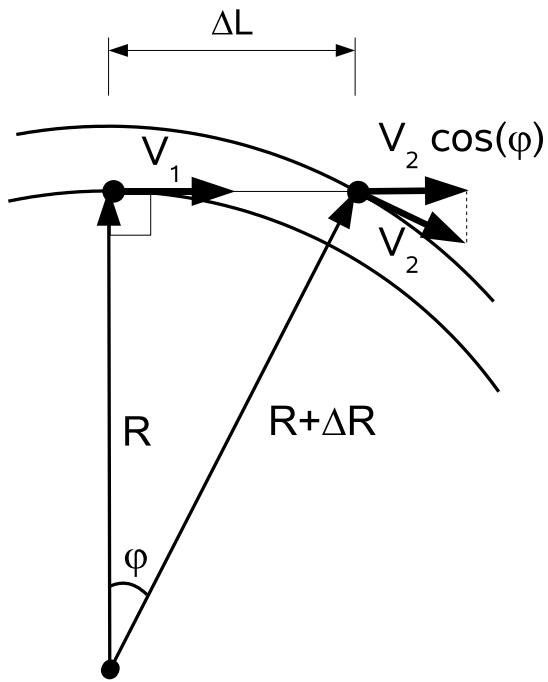}
\caption{Analytical estimation of the coherence length for a rotating disk.}
\label{scheme_LVG}
\end{figure}

The coherence length $\Delta L$ in the equatorial plane of the Keplerian
disk can be roughly estimated as follows (see Fig.~\ref{scheme_LVG}).
The photons emitted in a disk cell at a radius $R$ cannot escape in the
radial direction $\vec{R}$ since there is no velocity shift in this
direction. Instead, we assume that the photons can only escape
in the tangential direction. Ideally, one would need to apply spatial
averaging to get a correct value of the coherence length, but this would
significantly complicate the realization of the LVG method.

The approximate coherence length can be expressed via the local line width
and the Keplerian velocity. The velocity shift along the coherence path is,
by definition, equal to the local line width $\Delta V$:
\begin{equation}
\Delta V = |V_2\cos(\varphi)-V_1|,
\label{dV}
\end{equation}
where $V_1$ and $V_2$ are the Keplerian velocities at the distances of $R$
and $R+\Delta R$, respectively, and $\varphi$ is the corresponding angle
between the velocity vectors (Fig.~\ref{scheme_LVG}). Expanding the Keplerian
velocity into Taylor series, we find that $V(R+\Delta R) \approx
V(R) + \partial V/\partial R \Delta R$, while $\cos(\varphi)
\approx R/(R + \Delta R)$. This allows to derive the value of $\Delta R$
from Eq.~(\ref{dV}):
\begin{equation}
\frac{\Delta R}{R} = \frac{2}{3}\frac{\Delta V}{V}.
\end{equation}
Note that $\Delta R$ is the width of the ring-like radiatively coupled
region depicted in Fig.~\ref{interac}.

From Fig.~\ref{scheme_LVG} we also find that the coherence length $\Delta L$
can be related to the radial shift $\Delta R$ and radius $R$
as $\Delta R/\Delta L=\Delta L/R$, and thus
\begin{equation}
\Delta L = R\sqrt{\frac{2}{3}\frac{\Delta V}{V}}~~\propto R^{5/4}
\label{coherent}
\end{equation}
for a constant line width $\Delta V$.
This approach is adopted in our LVG method. It typically takes only
several seconds of CPU time to calculate the molecular level populations
with the LVG approach.

\subsection{Vertical escape probability}
\label{VEP}
One has to keep in mind that in our disk models there is no velocity gradient
in the vertical direction and the disk cells are radiatively coupled in
vertical direction, while in the LVG approach, the disk is effectively decoupled
in independent parallel planes. Let us compare the coherence length $\Delta L$ with
the scale height of the disk $H_p$. For the geometrically thin,
isothermal disk $H_p=\sqrt{2} C_s/\Omega$
\citep[for our definition of $H_p$, see][]{DDG03}, where $C_s$ is the sound speed and
$\Omega$ is the Keplerian angular velocity at radius $R$. For grey dust
opacities the radial temperature profile is $T \propto R^{-1/2}$ \citep{Dullemond_2002} and thus
$H_p \propto R^{5/4}$, similar to the scaling of the coherence length.
For our disk model $H_p\approx 28$~AU at $R$=100~AU,
while $\Delta L \approx 25$~AU. Therefore, the radiative effects in vertical
and radial directions are comparable over the entire Keplerian disk (at least around the
disk mid-plane) and, by design, the LVG method can only partly account for the radiative coupling.

\begin{figure}
\includegraphics[width=0.45\textwidth]{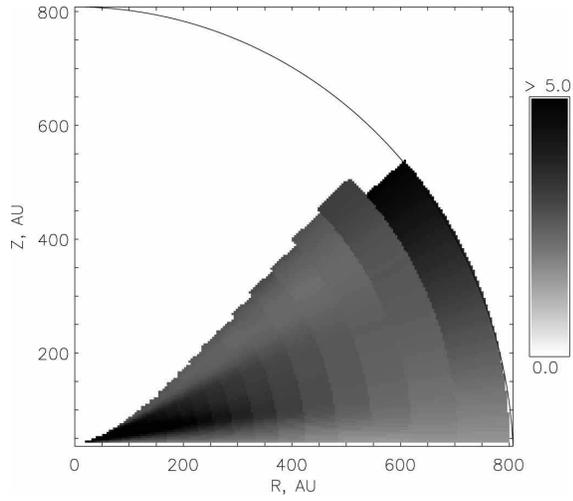}
\caption{Ratio of tangential to vertical optical depths calculated with ART
for the uniform disk model and the HCO$^+$(4-3) line.}
\label{tau_ratio}
\end{figure}

\begin{figure}
\includegraphics[width=0.45\textwidth]{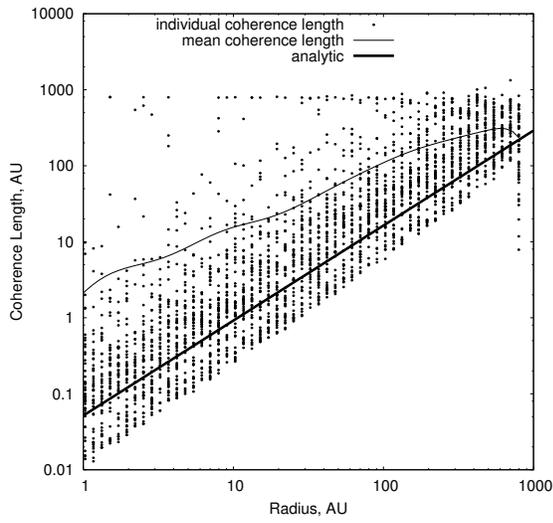}
\caption{Radial dependence of the coherence length (Eq.~\ref{coherent}; thick solid line)
compared to the individual (circles) and 3D spatially-averaged
(thin solid line) coherence lengths (ART). The uniform model and the HCO$^+$(4-3) line are used.}
\label{av_coh_length}
\end{figure}

To better illustrate the role of radiative coupling in vertical and radial directions,
we computed with ART the ratio between tangential and
vertical optical depths for the uniform disk model (see Fig.~\ref{tau_ratio}).
The typical value of this ratio is only about a few over the entire disk, so both
directions are important for the line radiative transfer.
Using ART, we also calculated the mean coherence
lengths in the equatorial plane of the disk. This is done by
spatial averaging of individual coherence lengths. The calculated mean and individual
coherence lengths are compared to the values from Eq.~(\ref{coherent}) in Fig.~\ref{av_coh_length}.
As can be clearly seen, our analytical expression reproduces well the radial dependence of the
mean coherence length, but a number of the individual lengths have larger values because
of the photon trapping along directions with zero Doppler shifts.
As a result, the mean value of the coherence length is typically about 10 times longer than
predicted by Eq.~(\ref{coherent}).

However, the excitation conditions vary so greatly over the mean coherence length that
it cannot be directly put into Eqs.~(\ref{eqlvg1}) and (\ref{eqlvg2}). In order
to properly evaluate the total escape probability $\beta$ one has to average
individual escape probabilities rather than the coherence lengths. This requires
information about exact physical parameters along all coupled directions, which
makes this approach non-local and thus far too complicated.

In the LVG approximation the coherence length is determined by the velocity
gradients resulting from the Keplerian rotation of the disk. However, as
we have shown above the radial radiative coupling is comparable to the
coupling in vertical direction. One can assume that the vertical radiative
coupling is more important than the radial coupling, which leads to
significant simplification of the radiative transfer.

This idea is utilized in the VEP method where photons are assumed to escape
only in the vertical direction perpendicular to the disk plane. A simple
correction is made to take into account the large asymmetry between the
two opposite escape directions: the escape probability is computed by taking
the sum of escape probability upwards and downwards:
\begin{equation}
\beta=\frac{1}{2}\left(\frac{1-\exp(-\tau^-)}{\tau^-}+
\frac{1-\exp(-\tau^+)}{\tau^+}\right),
\label{tauvep}
\end{equation}
where  $\tau^+$ is the optical depth towards the disk surface
(upward opacity), and $\tau^-$ is the optical depth through the disk
midplane (downward opacity). The optical depth in upward and downward
directions is calculated using the local excitation conditions and the
molecular surface densities along the corresponding directions. Note that
this assumption is rather coarse in the presence of strong vertical gradients
of physical conditions.

Similar to LVG, in the VEP approach the escape probability, source function,
and the local optical depth are obtained iteratively. However, the direct
implementation of the iterative VEP approach can be dangerous for disk
cells with negative excitation temperature. In VEP these local conditions
are extrapolated over the total vertical extent of the disk and the resulting
optical depth can be negative and large. This leads to the physically unrealistic
situation that the escape probability can become large, especially during the
iterations required to solve the statistical equilibrium equations and the VEP method
may fail to converge. To avoid this problem, we control the
downward line optical depth $\tau^-$ such that it cannot become negative.
The upward optical depth is allowed to remain negative, in order to
represent weak inversions at the surface layers.

The method was initially developed in the
DiskFit software package (see section 4.6).
Like FEP, the VEP method takes only a few seconds of CPU time to solve
the LRT problem (see Table~\ref{LRT_methods_table}).

\subsection{Non-local 1D-method VOR}
\label{VOR}
In the LTE and FEP methods radiative coupling effects are completely ignored,
while in the LVG and VEP approach radiative trapping is only partially treated.
Since there is no vertical velocity gradient in the adopted disk model, the
cells are generally radiatively coupled in vertical direction.
Furthermore, Fig.\ref{interac} shows that,
with the exception of a relatively narrow cone in the radial direction, a given cell in
the disk is only coupled with cells at the same radius, i.e. with regions with the
same 1D structure.

Using these ideas, we construct a non-local approximate LRT method in which the
radiative coupling in only the vertical direction is taken into account
(``VOR'' -- Vertical One Ray method). This simplifying assumption allows
to use a 1D approach for calculating the mean line intensity $J$ in a disk cell:
\begin{equation}
J = \frac{1}{2}\left(I_1+I_2\right),
\label{eqVOR}
\end{equation}
where $I_1$ and $I_2$ are the intensities of the radiation coming along vertical
direction from the top and bottom of the disk, respectively. To further accelerate
this method, only one central frequency for each emission line is taken into
account. This non-local RT problem is finally solved using only 2 photon paths
along vertical direction for the ray tracing. Computational demands of the
1D VOR method are much less than those of the exact 2D code ART. It usually
takes about a few minutes of CPU time to obtain the level populations.
Note that our VOR approach is another implementation of the Feautrier method
\citep[see][]{Feautrier1964} and resembles the 1D scheme of \citet{Lucas_1974}.

The 1D methods have been used to represent spheres as well as infinite planes. In the
latter case, an Eddington factor of 3 is used to take into account the mean opacity
over all possible angles. In our case, since the interaction only occurs in a ring
centered around the star, the Eddington factor should be smaller. We have simply used
the Eddington factor of 1, like for a sphere.

\subsection{DiskFit}
The described approximate methods are included in the ``URAN(IA)''
package. In our study we also used another software package ``DiskFit''.
The program DiskFit allows to model line radiative transfer in disks
using various approximate methods as well as to reconstruct basic disk
parameters using an advanced $\chi^2$-minimization scheme in the $uv$-plane
\citep{Dutrey_etal1994,GD98}. The program offers three possibilities
to solve the coupled equations of statistical equilibrium and RT: LTE, VEP,
and a non-local 1D approach. The latter is based on the Feautrier scheme
\citep{Feautrier1964} and is an adaptation to the disk geometry of the code
developed by \citet{Lucas_1974} for the line radiative transfer modeling
of interstellar clouds. This 1D method is thus similar to VOR, but uses
several frequency points across the line profile. The Eddington factor may also
be adjusted if needed.

Once the level populations are derived, the interferometric visibilities
as well as beam-convolved channel maps can be computed using the 3D ray tracing part
\citep{Dutrey_etal1994}. Any arbitrary orientations of the disk defined by
the inclination and position angles can be used. For that, a three dimensional
regular grid is used, with additional refinement for the central disk region
where the surface brightness gradient is very strong. The DiskFit radiative
transfer methods and the ray tracing part were thoroughly compared against
those implemented in the ``URAN(IA)'' package and good overall agreement was
achieved for the disk level populations and spectral maps for both the optically
thin and thick molecular transitions. Therefore, for the rest of
the paper we compare approximate and exact LRT methods from the ``URAN(IA)''
package only.

\section{Results}
\label{results}
Prior to the detailed comparison of various radiative transfer methods described above,
we discuss the problem of molecular line formation and excitation conditions in disks
in general.

\subsection{Molecular line formation in protoplanetary disks}
\label{line_form}

\subsubsection{Excitation regimes with uniform abundances}
\label{line_form_uni}
Since the thermal and density structure of protoplanetary disks is
in relatively well understood \citep[e.g.,][]{Dalessio_etal1998,
Dullemond_Turolla2000,Dalessio_etal2001}, we can discriminate
distinct regimes of line formation.

A particular rotational transition is excited when the molecular
hydrogen density exceeds a critical density $n_\mathrm{cr}$:
\begin{equation}
n_\mathrm{cr}=\frac{A_{ul}}{\sum_iC_{ui}}.
\end{equation}
Here $A_{ul}$ is the Einstein coefficient and $C_{ui}$ are the collisional
rates from the upper level $u$ to all other lower levels $i$.
The density, $n_\mathrm{th}$, at which the molecular levels become predominantly
populated by collisions and thus are thermalized is typically about one
order of magnitude higher than the critical density. We will further refer
to $n_\mathrm{th}$ as to the thermalization density.

Consequently, the disk can be roughly divided in two regions. In the so-called
``super-critical'' region near the disk midplane, where the density is larger
than the thermalization density, molecular level populations follow the Boltzmann
distribution and the excitation temperature is equal to the kinetic temperature,
$T_{\rm ex}=T_{\rm kin}$. In this region a simple LTE approach can be safely
utilized.

In the upper, more diffuse ``sub-critical'' region of the disk the line excitation
is partly controlled by collisions and partly by internal/external radiation.
The crucial parameter that determines the excitation conditions in this region
is the molecular column density. If this column density is so high that the
sub-critical region is opaque to the internal radiation, then the collisional
and radiative (de-)excitations are in perfect balance and the corresponding
molecular levels are thermalized, $T_{\rm ex}=T_{\rm kin}$. In contrast, when
the column density is so low that the sub-critical region is fully transparent
to the internal radiation, the corresponding molecular levels are mostly
radiatively populated and therefore subthermally excited:
$T_{\rm ex} \approx T_{\rm bg}$\footnote{Here $T_{\rm bg}$ is the
temperature of the background radiation, e.g. 2.73~K for the CMB field}.
In this case a simple Full Escape Probability approximation can be used
(see Section~\ref{results}). The characteristics of the emergent molecular
line spectrum will depend on the relative contribution from each of
the super-critical and sub-critical regions.

\subsubsection{Effects of chemical stratification}
This simple scheme becomes more complicated when the effects of chemical
stratification are taken into account. In this situation three distinct
regimes of line excitation can be distinguished.

\begin{figure}
\includegraphics[width=0.95\textwidth]{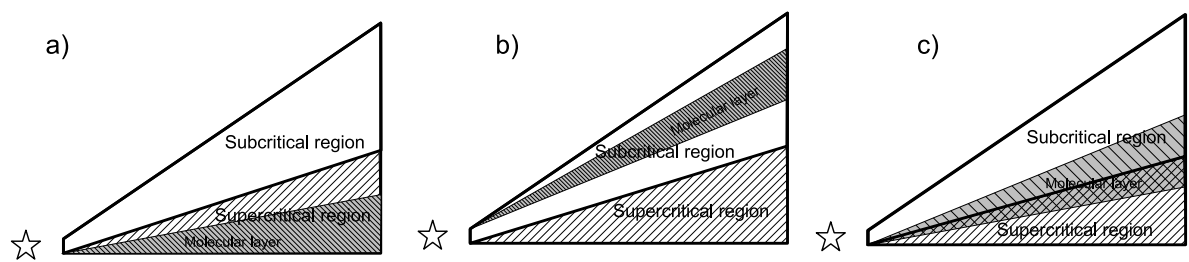}
\caption{Distinct cases of line formation in protoplanetary
disks with chemical stratification. Left: The molecular layer
is located inside the super-critical (high) density region
adjacent to the midplane and thus LTE holds for the level
populations. Middle: The molecular layer is located within
the sub-critical (low) density region close to the disk surface.
The level populations can be accurately calculated either with
the FEP method when the molecular
column density is sufficiently low or require a more sophisticated
non-local LRT method (e.g. VOR) when the molecular column density
is so high that radiative coupling within the disk becomes important.
Right: In the most general case the molecular layer lies both in
the sub- and super-critical regions. Consequently, FEP is only
accurate when the emission mostly comes from the super-critical
region of the molecular layer and the optical depth of the sub-critical
part is low. In all other situations non-local line radiative transfer
methods have to be applied.}
\label{scheme_LRT}
\end{figure}

In the first, and most simple case the molecular layer is located deeply
inside the super-critical density region and the molecular level populations
are in LTE (see left panel in Fig.~\ref{scheme_LRT}). Good examples of that
kind are the lower rotational transitions of CO and its
isotopologues \citep{DDG03}.

In the second case, the molecular layer is entirely confined within the
sub-critical density region. The distribution of the level populations will
then depend on the molecular column density as discussed above (middle panel,
Fig.~\ref{scheme_LRT}). When this column density is low and radiative coupling
is not important, the FEP approximation works. In the opposite case
the radiative transfer problem has to be tackled with a more sophisticated
non-local method. The line excitation proceeds in this way in gas-deficient
transition disks \citep{Jonkheid_etal2006}.

Finally, in the most general case the molecular layer is located between the
sub-critical and super-critical disk regions (right panel in Fig.~\ref{scheme_LRT}).
In this situation the LTE approach does not work, whereas the FEP
approximation is valid only if the optical depth of the sub-critical layer is
low and the molecular emission is dominated by the super-critical (thermalized)
layer. The lower rotational lines of e.g. HCO$^+$, HCN, and CS are likely excited
in such a way.

If the molecular column densities are so large that the sub-critical region
becomes optically thick for internal radiation, the simple FEP approach fails and
non-local radiative transfer modeling is required. The non-LTE effects
in disks are likely important for the rotational lines of complex species,
e.g. H$_2$CO, but may also be important for higher transitions of many other
simple molecules (such as HCN, CS, etc.). In the next section we verify whether
non-LTE effects are important in protoplanetary disks and for which molecular
transitions this is the case. There we use a simple preliminary analysis of the
excitation conditions.

\subsection{Critical column density diagrams}
\label{CCD_diags}
It is often useful to understand under what conditions a given transition
is excited before any LRT modeling is performed. One example is the
iterative $\chi^2$-minimization analysis of interferometric data where an approximate
LRT method has to be utilized due to limited computational resources
\citep[e.g.,][]{GD98,DDG03}.
Usually the basic properties of the physical structure and sometimes
the chemical structure of the disks are at least partly known. When the
chemical structure is not known, one can use uniform abundance distributions
with the abundance values taken from theoretical models.

Using the density and chemical structure of the disk and molecular data
as input, one can determine the line optical depth and the amounts of
emitting molecules in the sub- and super-critical regions. As discussed
in Section~\ref{VOR}, radiative coupling is strong in vertical direction
due to the absence of velocity gradients, and thus radiative effects
on the line excitation are most profound for face-on oriented disks.
Plotted as a function of radial distance, the molecular column densities
in the sub- and super-critical disk regions as well as the molecular column
density needed to reach an optical depth of unity may serve as an indicator
whether and where the considered transition is non-LTE excited and which
approximate or exact method is appropriate to model the radiative transfer
in this case (see previous section).

\begin{figure}
\includegraphics[width=0.6\textwidth]{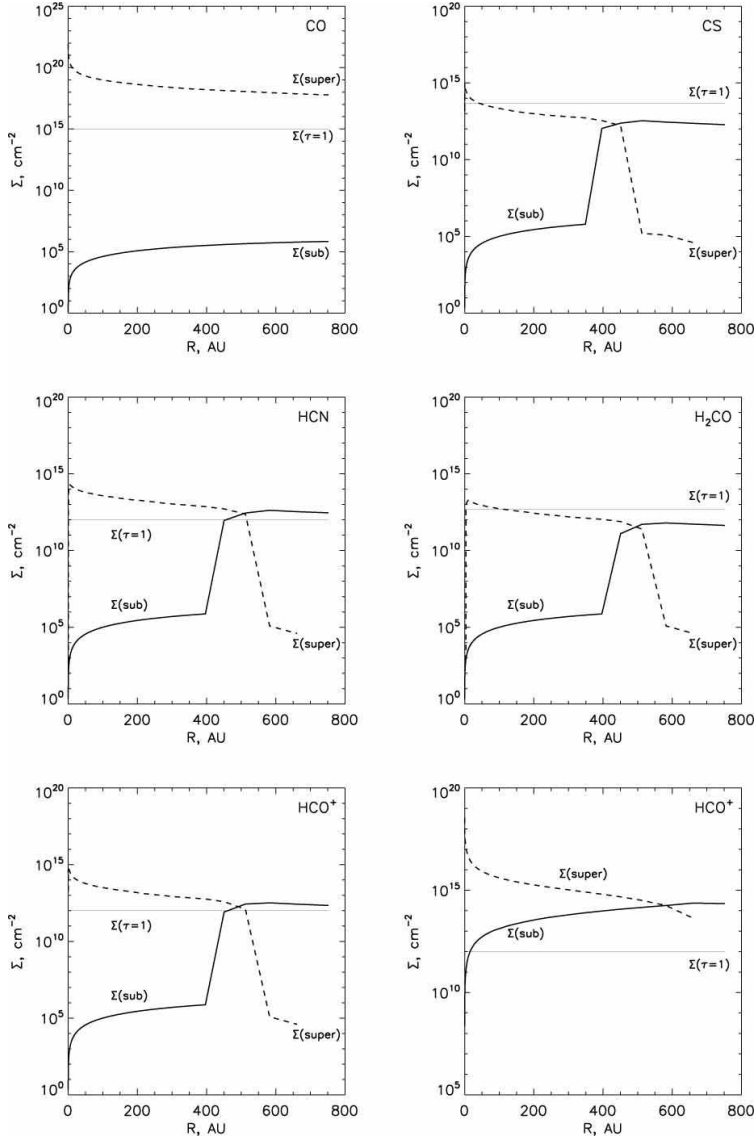}
\caption{(From top to bottom, from left to right)
Critical Column Density (CCD) diagrams for CO, CS, HCN, H$_2$CO,
and HCO$^+$. For each of these species, the CCD diagram shows the
radial distribution of 3 column densities in the chemically stratified
disk: the column density in the sub-critical layer (solid line), in the super-critical
layer (dashed line), and the column density that is needed to reach
an optical depth of unity for the (1-0) transition ($1_{01}-0_{00}$ for H$_2$CO)
($\tau\sim1$, thin solid line). The uniform HCO$^+$ model is shown in the lower right
panel. Note that for better presentation the column
densities are restricted to values above $\approx 10^6$~cm$^{-2}$.
The ratio between the column densities in the sub- and super-critical
layers, together with the value for $\tau=1$, can be used for
preliminary analysis of the line excitation conditions in disks
and stimulate the choice of a proper LRT method.}
\label{analysis_LRT}
\end{figure}

In Fig.~\ref{analysis_LRT} we present such Critical Column Density (CCD)
diagrams for some transitions of the molecules under investigation: CO, CS, HCN, H$_2$CO,
and HCO$^+$. We focus on low rotational transitions and adopt approximate
values of their thermalization densities, viz., $10^5$~cm$^{-3}$ for
the CO lines and $5\times 10^6$~cm$^{-3}$ for all other molecules. The
column densities to attain $\tau=1$ are computed for the central velocity
channel of the (1-0) transition ($1_{01}-0_{00}$ for para-H$_2$CO)
assuming a constant excitation temperature of 10~K.

Not surprisingly, the column density of CO is so high, $N(\mathrm{CO}) \sim 10^{18}$\,cm$^{-2}$,
that the lines are very optically thick,
$\tau\sim 10^2$--10$^3$. Since the CO lines are easily excited already
at densities of about 10$^3$\,cm$^{-3}$, essentially all CO molecules
are located within the super-critical region (case ``a)'' in Fig.~\ref{scheme_LRT}).
Molecules tracing the disk interior are excited in similar manner,
e.g. H$_2$D$^+$, N$_2$H$^+$, DCO$^+$ \citep{Aikawa_Herbst2001,Semenov_etal2004,CID1}.
Note that in our model, the CO abundances are high (see Table 2). When
comparing with real disks, where CO can be significantly depleted, our test case
for C$^{18}$O can be applied to $^{13}$CO observations, while for $^{12}$CO, the LTE
approximation will work even better (since $\tau=1$ will be reached at higher densities).

For other, less abundant species that have higher critical densities and are
concentrated in the disk intermediate layer, the CCD diagrams show a
bimodal behavior. In the inner, dense disk region at $r\la 400$--600\,AU
almost all molecules are located in the super-critical region,
and their levels are LTE-populated (similar to CO). In the outer, less dense
disk region the molecules are mainly concentrated in the sub-critical layer
(case ``b)'' in Fig.~\ref{scheme_LRT}). The optical thickness of the sub-critical
region for the disk model with chemical stratification is lower or close to one.
Consequently, the molecular sub-critical layers are expected to be optically
thin or moderately optically thick in the outer disk region and the FEP
approximation holds (see HCN and HCO$^+$ in Fig.~\ref{analysis_LRT}).
Thus, non-LTE effects should not play a major role for low rotational lines
of most molecules in protoplanetary disks. A notable exception is water that
has a complex level structure with a number of maser transitions
and some highly optically thick lines \citep{Poelman_etal2006,van_der_Tak_etal2006}.

In contrast to the layered chemical structure, for the uniform model the amount
of molecules is so high in the sub-critical region that this region is
optically thick and radiative trapping becomes efficient. As a representative
example, we present CCD diagrams for the uniform HCO$^+$ model (lower right
panel in Fig.~\ref{analysis_LRT}). In what follows, we will focus on this
uniform HCO$^+$ model to analyze in detail the importance of non-LTE
excitation for both low and high transitions as modeled by various LRT methods.

\subsection{Excitation temperatures}
\label{exc_temp}

\begin{figure}
\includegraphics[width=0.5\textwidth,clip=]{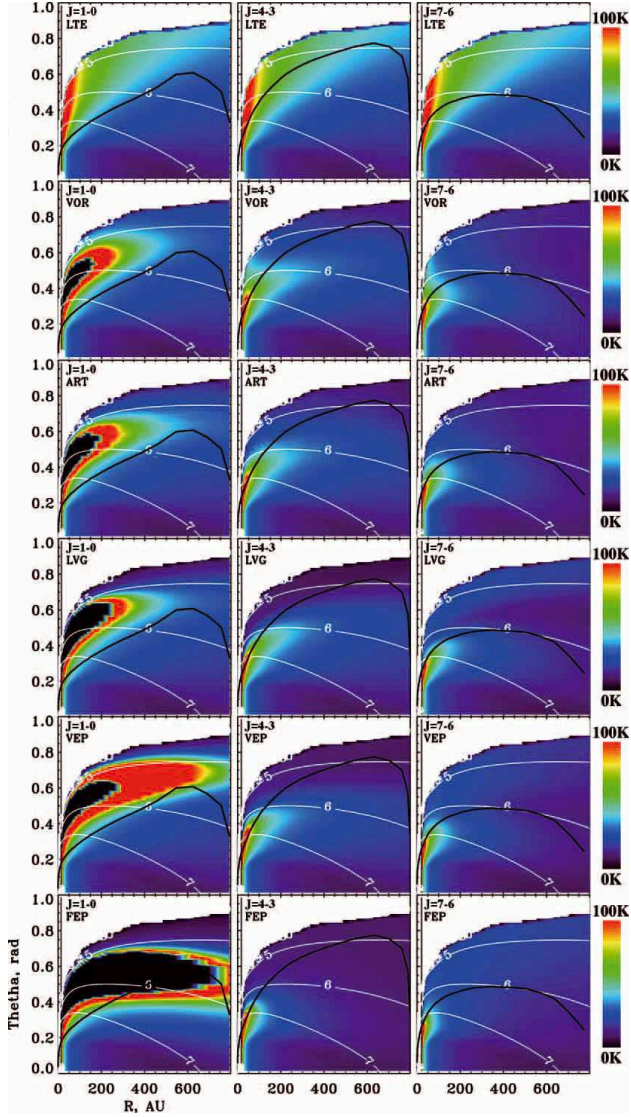}
\caption{(From left to right) Distributions of excitation
temperature (gray scale, in Kelvin) for the HCO$^+$ $J=$(1-0),
(4-3), and (7-6) transitions computed with the uniform disk
chemical structure (polar coordinates).
The black line depicts the location of $\tau=1$ for the corresponding
transition, as seen from above the disk in vertical direction.
Vertical scale is angle theta in radians. Shown are the results
for the LTE, VOR, ART, LVG, VEP and FEP radiative transfer methods
(from top to bottom). The white contours correspond to number
densities of $10^5$, $10^6$, and $10^7$~cm$^{-3}$. The white area
in the (1-0) distributions marks the zone of negative excitation
temperatures (maser effect).}
\label{Texc}
\end{figure}

In Fig.~\ref{Texc} we show the spatial distributions of the excitation
temperature in the uniform disk model as computed by the different LRT
methods for the HCO$^+$ (1-0), (4-3), and (7-6) transitions. By definition,
the LTE excitation temperature for any transition is equal to the gas
kinetic temperature. The vertical temperature gradient and cold isothermal
midplane are clearly visible (Fig.~\ref{Texc}, top row). In contrast,
the excitation temperature distributions calculated with other LRT methods
show a departure from LTE conditions, in particular for high transitions
and for the upper, less dense disk regions.

The most extreme deviation from LTE is the negative excitation temperature
caused by the inversion in the populations of the first two rotational
levels of HCO$^+$ (maser effect), which is reached in the warm upper disk
region, at $r\la 200$~AU (white zone in Fig.~\ref{Texc}, left panels).
In this part the (self-)radiation is not intense, and the levels are
excited and de-excited by collisions but only radiatively de-excited.
Consequently, LTE is not longer valid and collisional pumping of the
rotational level populations appears as a result of a specific ratio
between radiative and collisional (de)-excitation probabilities. Such
inversion is characteristic of some linear molecules like CS, HCO$^+$,
and SiO and requires high temperatures and a specific density range
\citep{LL77,VH78}.

However, this inversion does not result in maser amplification of the
HCO$^+$(1-0) line intensity even for highly inclined disks. As soon
as the line opacity and thus the self-radiation becomes
significant, the stimulated radiative transitions depopulate the
upper levels and destroy the inversion. Note that in the FEP
approximation the effect of self-radiation is completely neglected
and therefore the negative excitation temperature zone
is strongly overestimated (Fig.~\ref{Texc}, left bottom panel).

Another non-LTE zone corresponds to the subthermally excited
sub-critical disk region, which occupies a significant disk volume
for high HCO$^+$ transitions.

To quantify the difference between the excitation temperature
$T_\mathrm{ex}$ computed with an approximate and the exact ART methods,
we introduce the local error $\epsilon$ in each disk cell:
\begin{equation}
  \epsilon = \frac{T_\mathrm{ex}-T_\mathrm{ex}^{\rm ART}}{|T_\mathrm{ex}|+|T_\mathrm{ex}^{\rm ART}|}.
\label{local_error}
\end{equation}
A relative difference of $20\%$ in the excitation temperature
corresponds to $\epsilon \simeq 0.1$, and a $35\%$ difference leads
to $\epsilon \approx 0.2$.

Consequently, the global accuracy $Y$ of an approximate method over
the entire disk can be evaluated with the following expression:
\begin{equation}
  Y = \frac{1}{M}\sum_{\tau_j<1}|\epsilon_j|M_j,
\label{global_error}
\end{equation}
where the absolute values of the local errors $\epsilon_j$ are summed
over the optically thin disk cells and weighted by the number of molecules
$M_j$ in each cell. $M$ is a total number of molecules in the optically thin
region. The optical depth at the line central frequency, $\tau_j$, is
calculated in the vertical direction using the ART level populations.

This global accuracy criterion is designed to directly relate the
difference in the excitation temperatures to the difference that will
appear in the spectra. We
consider disk cells with $\tau \la 1$ (in vertical direction) because this is the
region where most of the line emission comes from (for face-on disks). In essence,
the errors in the $\tau \gg 1$ regions are not important, since these regions are not
visible. Certainly, this criterion is not a unique one and has to be used with care.

\begin{figure}
\includegraphics[width=0.5\textwidth,clip=]{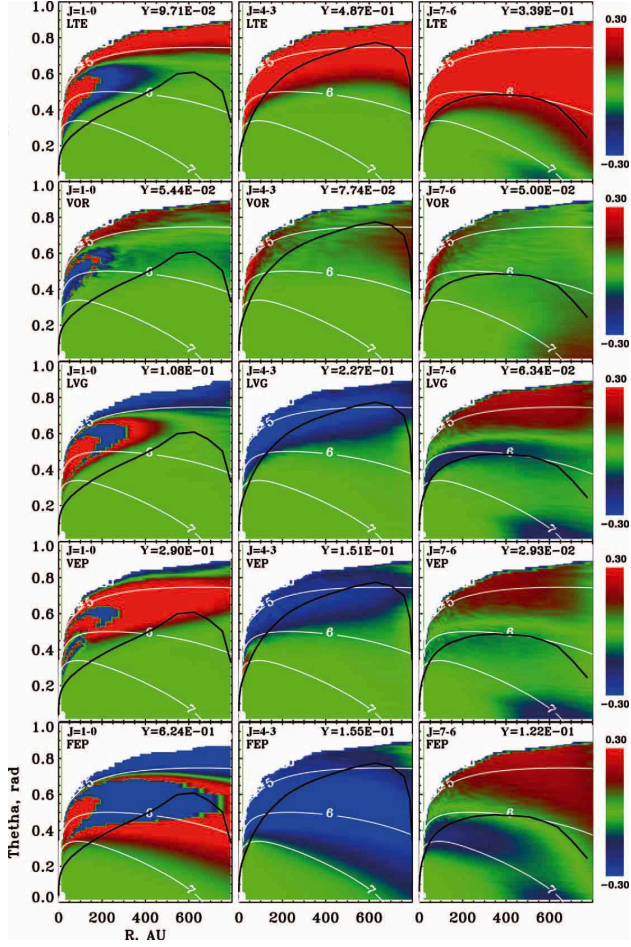}
\caption{The same as in Fig.~\ref{Texc} but for the local accuracy
estimate $\epsilon$. The thick black line depicts the location of $\tau=1$
for the corresponding transition, as seen from above the
disk in vertical direction. The accuracy of an approximate LRT
method is good when the global error $Y<0.1$, moderate when
$0.1<Y<0.2$, and bad when $Y>0.2$.}
\label{Texc_rel}
\end{figure}

We state that an approximate method gives
``good'' accuracy (``A'') if the global criterion $Y<0.1$, ``reasonable''
accuracy (``B'') when $0.1<Y<0.2$, and ``bad'' accuracy (``C'') when
$Y>0.2$. The distributions of the local error $\epsilon$ and global
accuracy $Y$ of the applied approximate LRT methods for the uniform
HCO$^+$ model are shown in Fig.~\ref{Texc_rel}.

All LRT methods are accurate in the super-critical disk region,
$n \ga 10^7$~cm$^{-3}$. However, the thermalized disk region extends
toward lower densities because the HCO$^+$ lines are highly optically thick
and the perfect balance between radiative and collisional (de-)populations
is reached (see Sect.~\ref{line_form_uni}). The thermalized region is
particularly large for low rotational transitions that are excited
at lower densities.

The LTE approximation tends to significantly overestimate the excitation
temperatures for the high transitions with high critical densities but
is still appropriate for the $J=$(1-0) line ($Y\la 0.1$, Fig.~\ref{Texc_rel}).
In contrast, FEP fails for the (1-0) transition since it greatly
overestimates the region of negative excitation temperature. It is rather
accurate ($Y\approx0.16$) for the (4-3) line because it adequately describes
the excitation conditions in the small, optically thin zone above
$\tau=1$ in the outer disk region, at $r\ga 400$~AU, which occupies the
largest volume in the disk. FEP is also accurate for (7-6) transition
($Y=0.12$), since it works in the large optically thin region that is
decoupled from the internal radiation.

The LVG, VEP and VOR methods are in general much more accurate than LTE
and FEP because they account for the radiative trapping. In particular,
the maser zone in the disk is well reproduced by LVG and VOR. The LVG
approach shows the same tendency for the excitation temperature deviations
as FEP because it underestimates the radiative coupling. The non-local
VOR approach gives the most accurate results for all considered
transitions since it treats the radiative transfer in vertical direction,
where velocity gradients are absent in our model. A minor problem of
VOR is the overestimation of the photon trapping in the very upper disk
regions where the radiation can escape also in other directions.

In general, for the uniform HCO$^+$ model and all considered transitions
FEP and LTE are not accurate LRT methods (``C'' grade), while LVG and
VEP are acceptable (``B''), and VOR is good (``A''). This conclusion
is valid in general for all other key molecules apart from CO isotopologues
where all methods are accurate because their lines arise
from the super-critical region (see Table~\ref{applic_methods_uni}).
For the disk model with chemical stratification the overall
accuracy of all LRT methods is better because the lines are more
optically thin and radiative effects are not that important
(Table~\ref{applic_methods_chem}). For many molecular transitions LTE
and FEP are reliable approximate methods for the line radiative transfer
modeling of chemically stratified disks. However, in this case both LVG
and VEP provide an even better approximation, at the expense of very
little extra computational time.


\subsection{Synthetic spectra and spectral maps}
\label{spectra}
\begin{figure}
\includegraphics[width=0.75\textwidth,clip=]{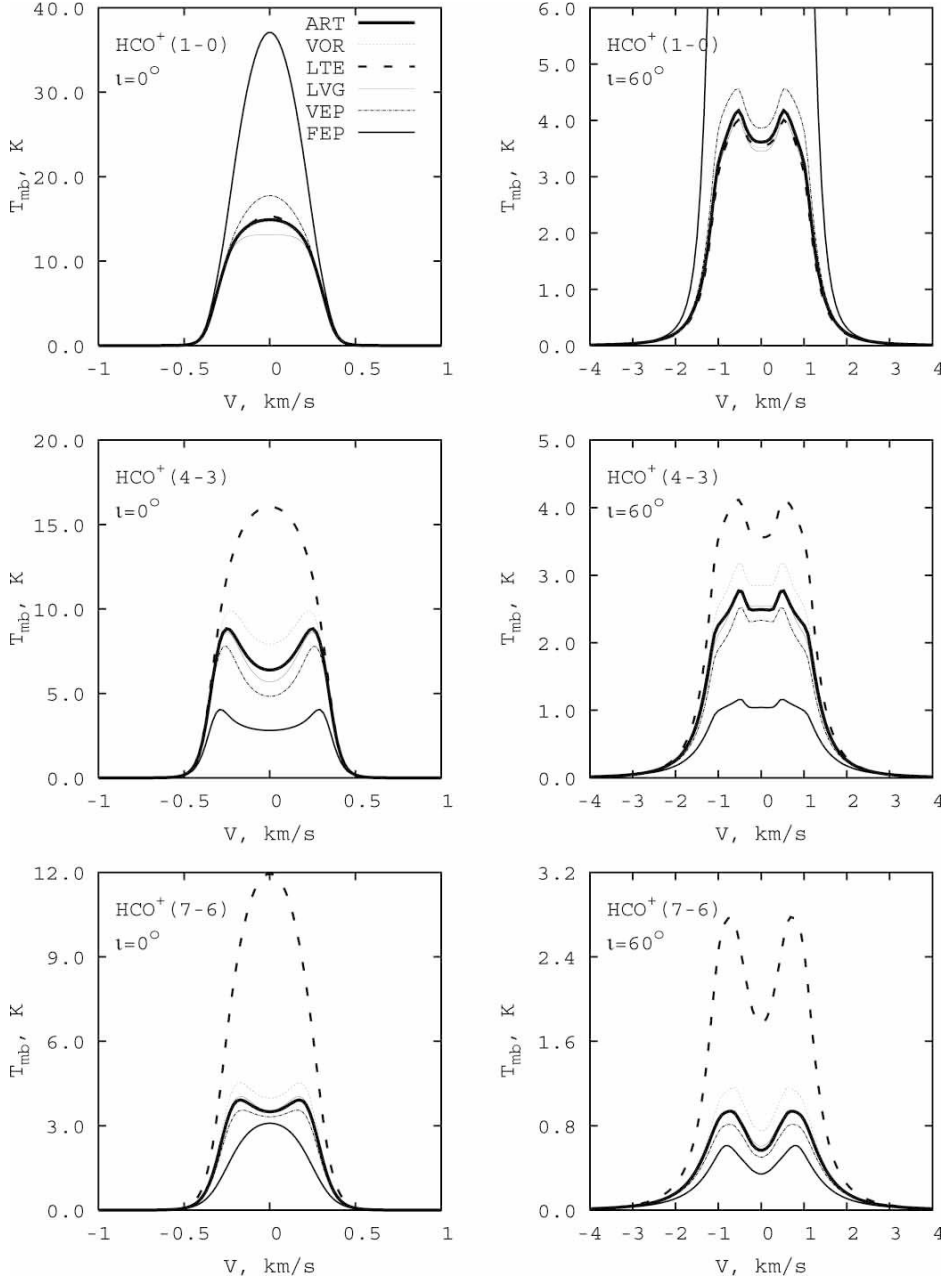}
\caption{(From top to bottom) Synthetic spectra for the HCO$^+$
(1-0), (4-3), and (7-6) transitions convolved with a $10\arcsec$
beam. The distance to the disk is 140~pc. The results are obtained
with the 2D exact ART (thick solid line), the 1D non-LTE VOR
(dotted line), LTE (dashed line), LVG (tiny solid line),
VEP (dash-dotted line) and FEP (thin solid line) approaches
using the disk model with uniform HCO$^+$ abundances. The disk
inclination is $0\degr$ (left panels) and $60\degr$ (right panels).
The intensity is given in units of the main beam temperature
(Kelvin).}
\label{single-dish_uniform}
\end{figure}

Using the excitation temperatures calculated for the uniform chemical
structure, we plot the corresponding synthetic HCO$^+$ spectra for the
$J=(1-0)$, (4-3), and (7-6) transitions in Fig.~\ref{single-dish_uniform}.
All spectra are convolved with the same $10\arcsec$ beam and are
computed for two disk inclinations: $i=0\degr$ (face-on) and $i=60\degr$).

The LVG, VEP and VOR methods result in line profiles with intensities deviating
by no more than 30\% from ART, whereas the FEP and LTE intensities and
line profiles differ more strongly from those of ART, in agreement with the difference in
excitation temperature maps (see Fig.~\ref{Texc_rel} and
Tables~\ref{applic_methods_uni},\ref{applic_methods_chem}). The line intensities computed with FEP
and LTE are either overestimated or underestimated, depending on
transition. The overall disagreement between these approximate methods
and ART does not depend strongly on the disk orientation.

For the HCO$^+$(1-0) transition the LTE spectrum is close to the ART line
profile, but the FEP intensity is 3 times higher for the face-on
orientation and even higher for the inclined disk. As discussed in the
previous section, this is due to the maser effect in the first two
HCO$^+$ level populations. The same effect applies to the lower
transitions of CS (not shown here). However, for the higher HCO$^+$
transitions the FEP spectra are in better agreement with the full Monte
Carlo simulations, whereas LTE tends to significantly overestimate the
line intensities. This is because the size of the radiatively
coupled part is decreasing for higher transitions, having higher
thermalization densities and thus a more extended sub-critical region
where $T_\mathrm{ex}\la T_\mathrm{kin}$ (compare top panels
in Fig.~\ref{Texc_rel}).

At face-on disk orientation a self-absorption dip arises in the (4-3)
and (7-6) ART spectra at the central velocity channels. This dip is caused
by absorption of emitted radiation in the upper disk region,
at $n\la 10^6$\,cm$^{-3}$, due to the vertical excitation temperature gradient
(Fig.~\ref{Texc}). Both LTE and FEP fail to accurately reproduce this effect.
In contrast, the double-peaked HCO$^+$ spectra of the inclined disk are not
due to self-absorption but appear as a result of the beam convolution over
disk cells with different projected velocity. In this case all approximate
methods correctly predict the shape of the line profiles. Note that while
integrated line intensities do not strongly depend on the disk orientation,
the peak intensities do differ strongly, and are lower for higher transitions.

\begin{figure}
\includegraphics[width=0.75\textwidth,clip=]{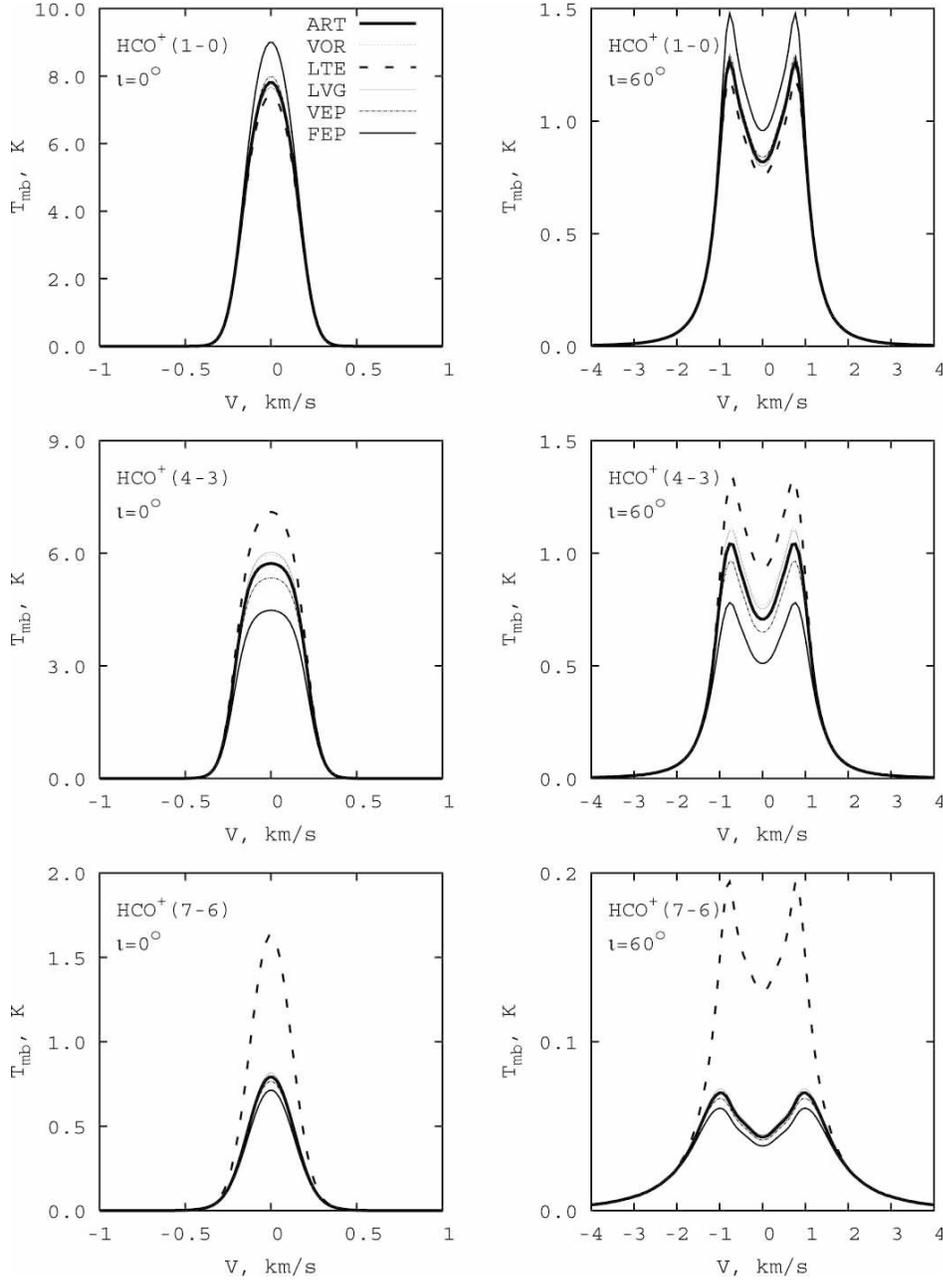}
\caption{The same as in Fig.~\ref{single-dish_uniform}
but for the layered HCO$^+$ abundances from the chemical
model.}
\label{single-dish_chem}
\end{figure}

The effects of chemical stratification on the integrated spectra are shown
in Fig.~\ref{single-dish_chem}. Since HCO$^+$ occupies only a fraction of
the entire disk, the computed line intensities and their optical depths
are much lower compared to the uniform model. The self-absorption dip is
absent in the face-on line profiles since there is no more HCO$^+$ in
the upper disk region. The overall agreement between the different LRT
methods becomes better (see also Table~\ref{applic_methods_chem}). Still,
LTE fails for higher transitions. The detailed comparison for the other
species considered here yields similar results.

\begin{figure}
\includegraphics[width=0.9\textwidth,clip=]{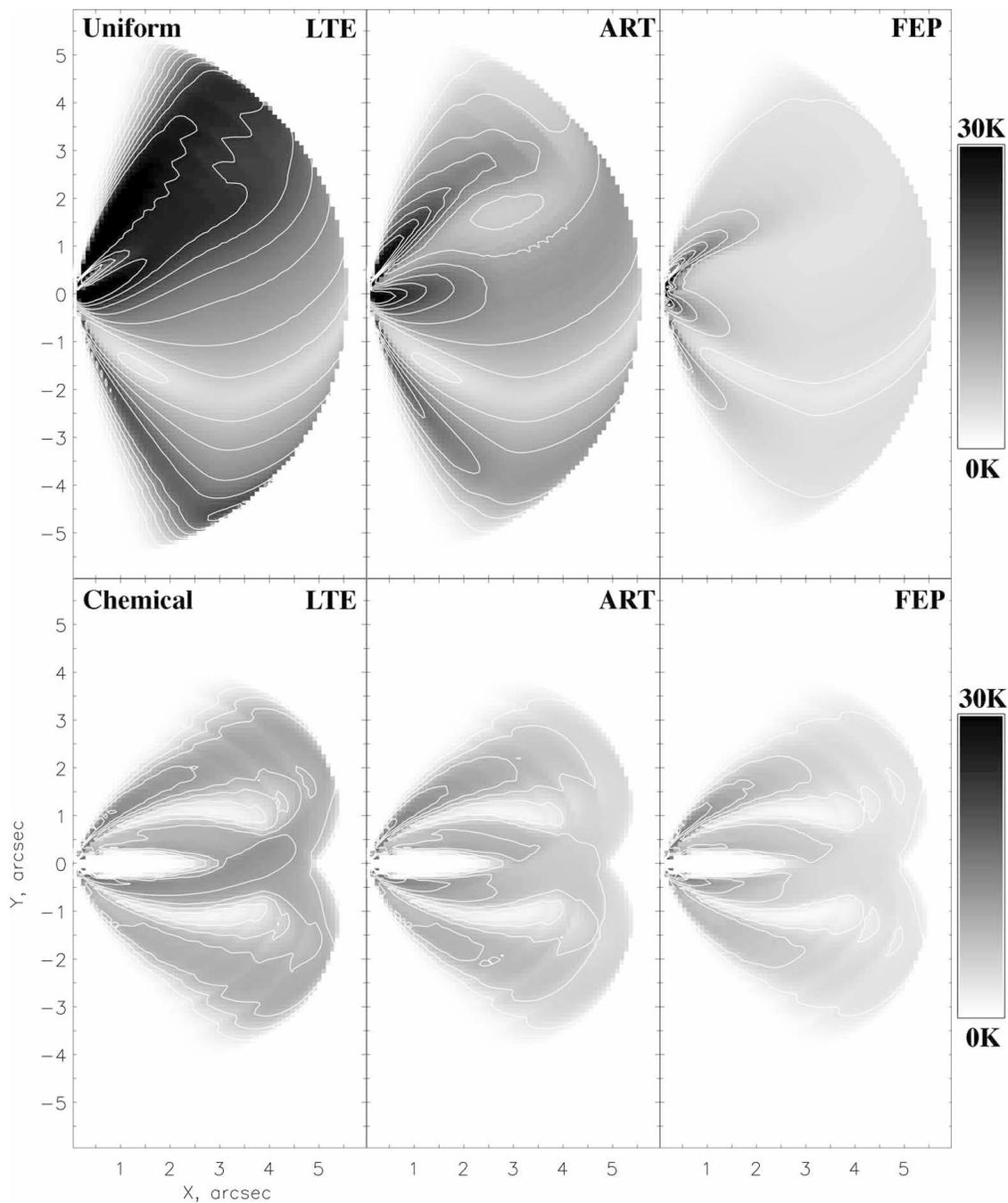}
\caption{HCO$^+$(4-3) intensity maps synthesized without beam
convolution for the 0.68~km\,s$^{-1}$ velocity channel and for
a disk inclination of $60\degr$. The distance is 140~pc.
The results are simulated using the level populations computed
with the LTE (left), ART (middle), and FEP (right panel) methods.
The uniform (top row) and layered (bottom row) abundances of
HCO$^+$ are utilized. The intensity is given in units of the
radiative temperature (Kelvin).}
\label{spectral_map}
\end{figure}

The criterion $Y$ that gives a global accuracy estimate for
the excitation temperature calculated with an approximate LRT method is also
valid for the single-dish spectra. To investigate how good this criterion is
for the synthetic spectral maps, we plot the HCO$^+$(4-3) map at the
0.68\,km\,s$^{-1}$ velocity channel for the same $60\degr$-inclined disk,
but without beam convolution (Fig.~\ref{spectral_map}).

The model with the uniform HCO$^+$ abundances leads to a fan-like structure
with two low-intensity lanes and a high-intensity interior zone
(see Fig.~\ref{spectral_map}, upper row, central panel). The low-intensity lanes
correspond to the cold disk midplane with low excitation temperatures.
Similar to the single-dish spectrum (Fig.~\ref{single-dish_uniform}),
LTE (FEP) tends to overestimate (underestimate) the total map intensity.
The LTE significantly overestimates the excitation temperature in the upper
disk region, leading to the formation of two high-intensity features near
the vertical edges of the image. The contrast of the FEP spectral image
is much smaller, yet the cold midplane and high-temperature inner zone are
clearly visible in the map.

The LTE, FEP, and ART HCO$^+$(4-3) maps simulated with the layered chemical
structure do not differ so much as for the uniform model
(see also Fig.~\ref{single-dish_chem}). However, the spectral images show a more
complicated pattern that resemble two thick elliptical rings overlaid
on each other with an angular shift of about $20\degr$ and fixed
at the center. The three low-intensity lanes correspond to the disk
region close to the midplane that is devoid of HCO$^+$ emission at the
chosen velocity channel of 0.68\,km\,s$^{-1}$. Furthermore, the HCO$^+$
ions are also absent in the disk upper layer and the overall extension of
the map is smaller in vertical direction compared to the uniform model.

\section{Discussion}
Our analysis of the line radiative transfer in protoplanetary disks
supports and extends the analysis performed by \citet{Zadelhoff_etal2001}.
In particular, they argued that SE($\bar{J}_{\nu}$=0)
calculations (in our notation FEP) provide better agreement with
SE calculations (in our notation ART) compared to LTE. According to
our results, the FEP method is indeed more correct than LTE for the
upper molecular transitions which were considered by
\citet{Zadelhoff_etal2001}. However, the FEP approach may give
completely wrong intensities for the lowest molecular transitions
such as $J=1-0$ and $J=2-1$ since it overestimates the maser effect,
resulting from the specific ratio between collisional
(de)excitation rates. Hence, the LTE approach provides more reliable
intensities of the lowest transitions than the FEP method.
However, LTE fails to reproduce important features of the line profiles
like the self-absorption dips.  Despite the fact that either LTE or FEP
can be sufficient for the simulations of the particular molecular lines,
we recommend to use LVG or VEP since they provide in general better
agreement at the same cost of complexity and CPU requirement. The more
exact, but more time-consuming methods such as VOR or ART can be used to
check the validity of the above approximate methods.

An interesting result of our study is the fact that the synthetic images calculated with
VEP are quite close to the exact ART spectra, for both chemically uniform and
layered models. This may look surprising since the idea of the VEP method is
to use local excitation conditions and to extrapolate them along the vertical
direction to calculate escape probabilities. Thus, one may expect that this
``homogeneous'' approach is inappropriate for the disk models
with their strong gradients of physical conditions and molecular abundances.

The ``good behavior'' of VEP is a consequence of the very strong (vertical) density gradient
compared to the temperature gradient. In the layered disk model molecules are
concentrated in thin layers where physical conditions do not change much in vertical
direction. In the uniform disk model the strong vertical density gradients result in a
super-critical region around the midplane where molecular emission is thermalized, while
emission from a low-density surface layer is negligible.
In between lies a narrow sub-critical disk region, where molecular opacities can
be high enough to affect the emergent spectra (see Sect.~\ref{line_form}).
For optically thin lines, the contribution of this narrow layer to the emission is small.
For optically thick lines, the likelihood that the line opacity reaches about unity in
this region is very low. The exact location of the $\tau=1$ layer determines the
brightness temperature, but is not very critical since vertical temperature gradients are
not very large.

Recently \citet{PS05} have proposed a new approximate Multi-Zone Escape
Probability (MEP) method to model water line emission from a PDR region \citep{PS06}.
In contrast to our VEP approach, in MEP the probability for photons to escape a
cell is computed in 3D, by spatial averaging of the optical depths over several (6)
directions. Due to its 3D nature, MEP may treat more consistently the photon escape from
complicated radiatively coupled regions in Keplerian disks (see Fig.~\ref{interac})
than our Vertical Escape Probability method. Since the 1D VEP approach already reasonably
accurate in most cases, the use of 3D MEP for the LRT modeling of protoplanetary disks
is recommended in situations when even higher accuracy is required at the expense of
extra CPU costs for the spatial averaging.

Another recent 1D escape probability method has been derived by \citet{EAR06}.
Their Coupled Escape Probability (CEP) approach is based on mathematical
transformation of the coupled system of the integro-differential RT equation and linear
balance equations into one system of implicit non-linear equations. Due to fast convergence
of this implicit scheme, CEP gains about 1--2 orders of magnitude in computational speed
compared to conventional ALI schemes. However, this method cannot be easily adapted to
multidimensional LRT problems with multi-level molecules because of the mathematical ``trick''
that lies behind its algorithm. Taken at face value, in 1D geometry, it would be conceptually
similar to, but more sophisticated than the 1D Feautrier approach implemented in our VOR method.

In the present study we did not consider molecules with complex
level structure, e.g. 
molecules with hyperfine components since they require more detailed investigations.
For instance, \citet{Daniel_ea06} used new collisional rates to
simulate the N$_2$H$^+$ lines from prestellar cores and showed
that the usual assumption that N$_2$H$^+$ sub-level populations follow the
Boltzmann distribution may lead to significant errors in the
intensities of the hyperfine lines. \citet{van_der_Tak_etal2006}
mentioned severe problems in LRT simulations of H$_2$O
lines. Apparently, the formation of such complex molecular lines
in protoplanetary disks requires a more extended analysis which has to be
based on the appropriate collisional rates.

In the present paper, we also did not treat the effects of dust emission on the line
formation. For the protoplanetary disks the dust emission at millimeter
wavelengths is negligible compared to the molecular transitions.
However, at shorter wavelengths the dust emission may be sufficiently
strong to excite upper molecular transitions and, therefore, should
be considered.

Nor did we include the effects of vertical mixing and turbulence in our model. From
a chemical point of view, turbulent diffusion acts as an efficient non-thermal desorption
mechanism and smoothes molecular abundance gradients in disks
\citep{Ilgner_etal2004,Semenov_etal2006,Willacy_etal2006,TG07}. Consequently, the results
for such a disk model would be in between the results for our uniform and layered chemical models.

\section{Conclusions}
\label{concl}
We analyzed the line radiative transfer and the formation of rotational
lines of several key molecules (CO, CS, HCO$^+$, H$_2$CO, HCN) in
protoplanetary disks. A detailed model of the disk physical structure
with vertical temperature gradient and uniform/layered chemical structure
is used. We used the exact Accelerated Monte Carlo code ART and
compare the results with several approximate LRT approaches,
including LTE, the Full Escape Probability (FEP), LVG, and
the Vertical Escape Probability method (VEP) and a 1D non-local code VOR.
Our main conclusions are as follows:

\begin{enumerate}

\item Prior to any modeling of the radiative transfer, one can
roughly estimate the importance of non-LTE effects for the line
excitation and thus choose an appropriate approximate or exact
method by using Critical Column Density (CCD) diagrams. The basic
idea of the CCD diagrams is to calculate and compare the amount
of emitting molecules in the sub- and super-critical regions,
taking into account the optical depth of the line. With CCD diagrams
we found that the upper disk layers are subthermally excited for many
molecular lines, in particular for the upper transitions.

\item By comparing excitation temperatures, synthesized spectra, and
spectral maps we demonstrate that the simple LTE and FEP methods work
reasonably well for chemically stratified disks where molecules are
located mainly in the warm intermediate layer or in the disk midplane, but
are not always accurate for disk models with uniform abundances.
In contrast, the LVG, VEP, and especially VOR methods take the radiative
coupling in the disks into account and thus accurately reproduce the
line intensities and profiles. Since the LVG and VEP methods are computationally
inexpensive, their use is recommended.

\item We found a possibility to accelerate the ray-tracing part
of the RT solver in Monte-Carlo method for rotating disks.
In this modification the ray tracing is only performed over radiatively
coupled disk zones, leading to a computational speed gain by factors of 10--50.
\end{enumerate}

This work is only the first step in our systematic analysis
of the line radiative transfer in protoplanetary disks. The
excitation of molecules with complex level structure, including
hyperfine line components, and the applicability of various
approximate LRT methods for the $\chi^2$-minimization fitting
of the observational data will be addressed in future studies.

\begin{acknowledgements}
We thank Michiel Hogerheijde and Jinhua He for useful and stimulating
discussions. Authors are thankful to the anonymous referee for valuable comments and
suggestions.
This research has made use of NASA's Astrophysics Data System.
\end{acknowledgements}


\clearpage

\clearpage

\clearpage

\clearpage

\clearpage

\clearpage

\clearpage

\clearpage

\clearpage

\clearpage

\clearpage

\clearpage

\clearpage
\begin{deluxetable}{lll}
\tablewidth{0pt}
\tablecaption{Adopted parameters of the disk and central star\label{disk_param}}
\tablehead{\colhead{Parameter} & \colhead{Symbol} & \colhead{Value} }
\startdata
Distance & $r_*$ & 140~pc \\
Stellar temperature & $T_\mathrm{eff}$ & $4\,000$~K \\
Stellar radius & $R_*$ & $2.64\,R_{\sun}$  \\
Stellar mass & $M_*$ & $0.7\,M_{\sun}$ \\
Disk inner radius & $R_\mathrm{in}$  & 0.037~AU\\
Disk outer radius & $R_\mathrm{out}$ & 800~AU\\
Disk mass & $M_\mathrm{disk}$ & $0.07\,M_{\sun}$ \\
Surface density at 1~AU & $\Sigma_1$ & $\approx 63^{*}$~g\,cm$^{-2}$ \\
Density profile & $p$ & $\approx -1^{*}$\\
\enddata
\tablenotemark{*}\tablenotetext{*}{Density structure is outcome of disk model.}
\end{deluxetable}

\clearpage
\begin{deluxetable}{lcccccc}
\tablewidth{0pt}
\tablecaption{Parameters of the disk chemical model\label{chem_param}}
\tablehead{
\colhead{Molecule} & \multicolumn{4}{c}{Molecular layer} & \colhead{Abundance,} & \colhead{Abundance,}\\
\colhead{} & \multicolumn{2}{c}{lower bound} & \multicolumn{2}{c}{upper bound}&
\multicolumn{2}{c}{$N(\mathrm{X})/N(\mathrm{H})$}  \\
\colhead{} & \colhead{$a_\mathrm{min}$} & \colhead{$b_\mathrm{min}$} & \colhead{$a_\mathrm{max}$} &
\colhead{$b_\mathrm{max}$} & \colhead{(layered)}  & \colhead{(uniform)}}
\startdata
CO         & 0.7  &  0.0   & 2.0  &  0.0   &  8\,(-05)$^a$  & 1\,(-04)\\
C$^{18}$O  & 0.7  &  0.0   & 2.0  &  0.0   &  2\,(-07)      & 2\,(-07)\\
HCO$^+$    & 1.37 & -0.1   & 1.68 & -0.05  &  6\,(-10)      & 1\,(-08)\\
DCO$^+$    & 0.0  &  0.0    & 0.4  &  0.0   &  5\,(-11)$^b$  & 2\,(-10) \\
H$_2$CO    & 1.37 & -0.1   & 1.2  &  0.0   &  3\,(-10)      & 1\,(-08)\\
HCN        & 1.75 & -0.16  & 1.92 & -0.07  &  1\,(-09)      & 1\,(-09)\\
CS         & 1.44 & -0.07  & 1.83 & -0.03  &  1\,(-09)      & 1\,(-08)\\
\enddata
\tablenotemark{a}\tablenotetext{a}{The value $A(-B)$ reads as $A\times 10^{-B}$.}
\tablenotemark{b}\tablenotetext{b}{DCO$^+$ is located in cold outer disk regions
at $r\ga100$~AU.}
\end{deluxetable}

\clearpage
\begin{deluxetable}{lclc}
\tablewidth{0pt}
\tablecaption{Overview of the applied LRT approaches\label{LRT_methods_table}}
\tablehead{\colhead{Method} & \colhead{Type} & \colhead{Applicability} & \colhead{CPU time$^*$}}
\startdata
LTE     & Local     & Super-critical         &  $<1$~s \\
        &           & region                 &          \\
FEP     & Local     & As above \& optically  &  $<1$~s \\
        &           & thin lines             &          \\
LVG     & Local     & Regions with strong    &  $\sim 1$~s \\
        &           & velocity gradients     &          \\
VEP     & Local     & Regions with low       &  $\sim 1$~s \\
        &           & velocity gradients     &          \\
VOR     & Non-local & Regions with low       &  $15$~m \\
        &           & velocity gradients     &          \\
ART     & Exact     & Everywhere in disks    &  $24$~h \\
\enddata
\tablenotemark{*}\tablenotetext{*}{CPU times are given for the $128\times128$ disk grid,
200 photons, and a Pentium~4 2.4~GHz PC.}
\end{deluxetable}

\clearpage
\begin{deluxetable}{llllll}
\tablewidth{0pt} \tablecaption{Reliability of the approximate LRT
methods for uniform model\label{applic_methods_uni}}
\tablehead{\colhead{Molecule} & \multicolumn{5}{c}{Methods} \\
\colhead{}  & \colhead{LTE} & \colhead{VOR} & \colhead{LVG} & \colhead{VEP}& \colhead{FEP}}
\startdata
CO          & AAB$^*$         & AAA             & AAA             & AAA            & ABB \\
CS          & ACC             & AAA             & ABA             & ABA            & BCB \\
HCO$^+$     & ACC             & AAA             & BAA             & CBA            & CCB \\
HCN         & ACC             & AAA             & ABA             & BAA            & CCB \\
H$_2$CO     & ACC             & AAA             & AAA             & AAA            & BCA \\
C$^{18}$O   & AAA             & AAA             & AAA             & AAA            & AAA \\
DCO$^+$     & ABC             & AAA             & AAA             & AAA            & AAA \\
\enddata
\tablenotemark{*}\tablenotetext{*}{``A'' - good accuracy
($Y\le0.1$); ``B'' - moderate accuracy ($0.1<Y\le0.2$);
``C'' - bad accuracy ($Y>0.2$). For all species except H$_2$CO the 1st, 2nd, and 3rd letters correspond
to the (1-0), (4-3), and (7-6) transition, respectively. For H$_2$CO the $1_{01}$-$0_{00}$,
$4_{04}$-$3_{03}$, and $4_{23}$-$3_{22}$ para-transition are considered.}
\end{deluxetable}

\clearpage
\begin{deluxetable}{llllll}
\tablewidth{0pt} \tablecaption{Reliability of the approximate LRT
methods for chemical model\label{applic_methods_chem}}
\tablehead{\colhead{Molecule} & \multicolumn{5}{c}{Methods} \\
\colhead{}  & \colhead{LTE} & \colhead{VOR} & \colhead{LVG} & \colhead{VEP}& \colhead{FEP}}
\startdata
CO          & AAA$^*$         & AAA             & AAA             & AAA            & AAA \\
CS          & AAC             & AAA             & AAA             & AAA            & AAA \\
HCO$^+$     & ABC             & AAA             & AAA             & AAA            & BBA \\
HCN         & ACC             & AAA             & AAA             & AAA            & CBA \\
H$_2$CO     & BBC             & AAA             & AAA             & AAA            & AAA \\
C$^{18}$O   & AAA             & AAA             & AAA             & AAA            & AAA \\
DCO$^+$     & AAA             & AAA             & AAA             & AAA            & AAA \\
\enddata
\tablenotemark{*}\tablenotetext{*}{See notation in Table~\ref{applic_methods_uni}.}
\end{deluxetable}

\end{document}